# Using a smart phone for information rendering in Computer-Aided Surgery

[1]Gaël Le Bellego, [1]Marek Bucki, [1,2]Ivan Bricault, [1]Jocelyne Troccaz[*]

[1]UJF- Grenoble 1 / CNRS / TIMC-IMAG UMR 5525, Grenoble, F-38041, France
[2]CHU Grenoble, Radiology and imaging Department, Grenoble, F-38043, France

first-name.last-name@imag.fr

**Abstract.** Computer-aided surgery intensively uses the concept of navigation: after having collected CT data from a patient and transferred them to the operating room coordinate system, the surgical instrument (a puncture needle for instance) is localized and its position is visualized with respect to the patient organs which are not directly visible. This approach is very similar to the GPS paradigm. Traditionally, three orthogonal slices in the patient data are presented on a distant screen. Sometimes a 3D representation is also added. In this study we evaluated the potential of adding a smart phone as a man-machine interaction device. Different experiments involving operators puncturing a phantom are reported in this paper.

## 1      Introduction

### 1.1    Computer-Aided Surgery (CAS) principles

For more than two decades, navigation systems are proposed to the clinicians to assist them during an intervention [1]. Typically CT anatomical data are collected for a given patient before the intervention. These data allow planning the intervention, for instance by defining a target position for a surgical instrument or for a prosthesis element. During the intervention, the navigation system gives information to the clinician about the progress of the intervention: typically the position of the instrument relatively to the target and to pre-recorded anatomical data is visualized in real-time. The position of surgical instruments in space is known thanks to a tracking device called "localizer." Most often, an additional stage is required to bring the surgical plan recorded pre-operatively to the intra-operative conditions; this stage is named registration. The approach is very similar to navigation assistance of cars. The localizer is similar to the GPS positioning system and the CT data correspond to the recorded road and city maps on which the position of the car is displayed.

---

[*] Author for correspondence : jocelyne.troccaz@imag.fr

Such navigation systems are also called "passive" assistance systems [2] since they only render information to the clinician who can use it in the way he/she wants. Alternative "active" assistance systems exist: in this case a robot can perform autonomously a part of the intervention. Intermediate "semi-active" solutions also exist where a robot may be tele-operated by the clinician or a programmable mechanical guide may constrain the possible motion of the instrument moved by the clinician. This paper focuses on passive navigation systems.

### 1.2   Man-machine interaction in CAS

One very conventional way of displaying guidance information to the clinician is based on the dynamic visualization of orthogonal slices computed in the volume of recorded data (see figure 1). A sagittal slice is a vertical slice which passes from front to rear dividing the body into right and left sections. The transverse slice (also called the horizontal slice or axial slice) is obtained by cutting the volume by a plane that divides the body into superior and inferior parts. It is perpendicular to sagittal slices. A coronal slice (also named frontal slice) is a vertical slice that divides the body into ventral and dorsal section. The intersection point of the three slices generally corresponds to the tip of the navigated instrument. When the instrument is moved the three slices are updated accordingly.

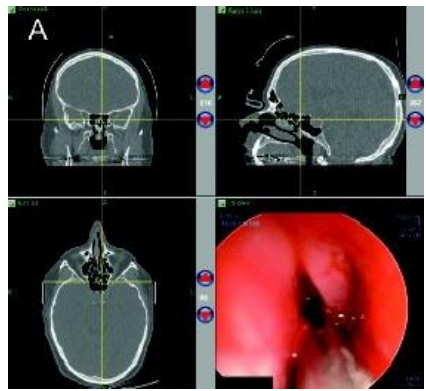

**Fig. 1.** Typical display of a navigation system. Coronal slice (top left), sagittal slice (top right), transverse slice (bottom left), endoscopic view. The position of the surgical instrument tip is visualized by the intersecting yellow crosses on each of the three slices. (Source: Neurosurgery Focus © 2008 American Association of Neurological Surgeons)

Because a meaningful representation of the tool trajectory is generally very important, the standard cutting planes presented above can be replaced by what is called pseudo-slices. A pseudo-transverse slice includes the tool axis and is slightly angulated with respect to a conventional transverse slice (see figure 2 left). Figure 2 right shows the GUI (Graphical User Interface) of a navigation system for punctures. A pseudo-transverse slice and a pseudo-sagittal slice help visualize the position and orientation of the puncture needle with respect to the patient anatomy.

When a target trajectory has been predefined in a planning stage, some additional information may be presented to the user, in order to compare the executed trajectory with the planned one. The trajectories can be visualized using specific visor displays in addition to the slice viewer. [3],[4] and [5] propose such "targeting" interfaces in their GUI.

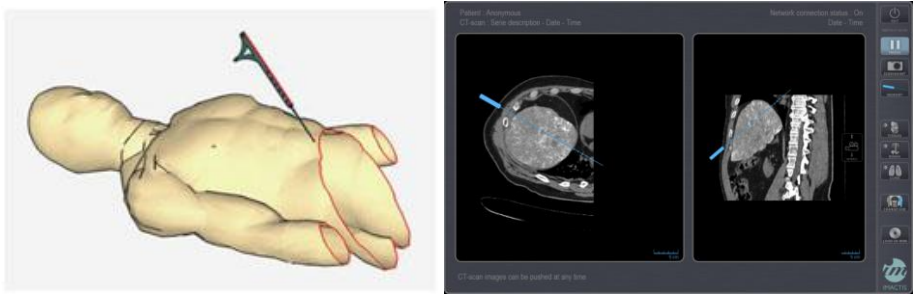

**Fig. 2.** Pseudo-slices. Definition of the pseudo-transverse slice (left). Two slices (pseudo-transverse and pseudo-sagittal) in a computer-assisted needle puncture software (right).

Most often data are displayed on a 2D screen installed in the viewing space of the clinician. Using the displayed information generally requires moving the surgical instrument without looking at it or at the patient. Perceptual continuity [6] is no longer guaranteed. Augmented reality systems have been proposed to remove this limitation. They are based on semi-transparent devices such as Head Mounted Displays [7] where navigation data is overlaid on intra-operative images given by an already existing sensor (for instance a surgical microscope [8]). Except for this last case, very few augmented reality systems are used in routine clinical practice.

More recently several groups [9], [10] proposed to display part of the guidance information on small mobile screens where the displayed data may depend on the position of the screen. [11] and [5] proposed to attach the screen to the instrument. [5] presents some experimental evaluation of different display modes.

A few years ago, thanks to the technology evolution in particular regarding PDAs and smart phones, our team decided to explore this potentially new interaction paradigm for CAS applications. The purpose of this work was to study the feasibility of using a mini-screen, within a close range to the operating site, in order to display partly or totally the guidance information to the clinician during interventions. Different combinations of displays and different representations of data and interaction modes with the data were explored for interventions such as punctures. The experimental environment and the conducted experiments are presented and discussed in the following sections.

## 2      Material and methods

### 2.1    Experimental environment

The system (cf. figure 3) includes the standard elements of a navigation system: the optical localizer (passive Polaris from NDI, Inc.) enables tracking in real time objects equipped with reflecting markers. It is linked to the computer running the navigation application. A 19 inches screen is connected to the computer; in the following we will call it the master screen. As regards the mini-screen, several possibilities were envisioned (LED, OLED, LCD screens, PDA, smart phones). We selected the iPhone3G which advantages were to have many embedded features (good quality display, wi-fi communication, accelerometers, tactile interaction, camera, microphone, etc.), a complete development environment and a large interest and experience from the HCI community. A client-server application controls the dialog between the main computer and the smart phone.

The user can interact with the navigation application on the main computer and master screen in a traditional way (scrolling, mouse clicking, etc.). When using the smart phone the interaction with the data is possible using the tactile screen (scrolling for zooming functions, clicking for definition and recording of a position of interest) and the accelerometers (for navigation around a point of interest).

The experiments are performed using a custom-made phantom. A block of deformable PVC in which the punctures are performed is placed inside a manikin. The puncturing instrument and the manikin are equipped with reflecting markers and are tracked by the localizer. CT data are associated to the phantom for navigation.

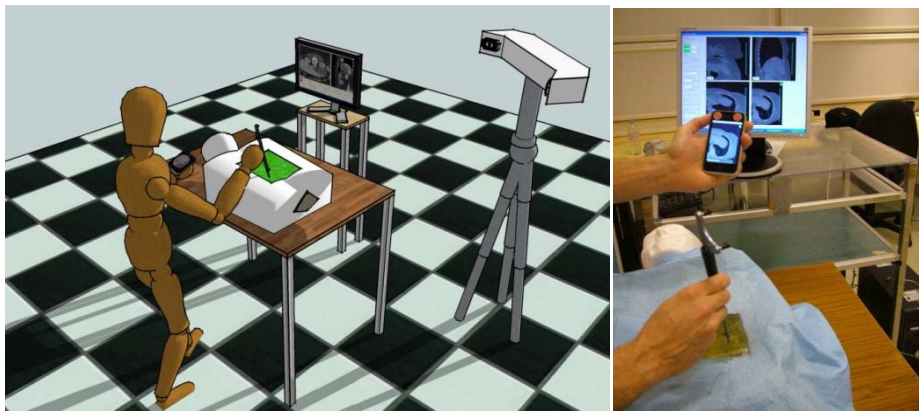

**Fig. 3.** Experimental set-up.

### 2.2    Data representation and operating modes

After having experienced several possible representations of data with users, we selected three of them:

- "Triple-ortho" representation: this is the standard representation used for navigation; three orthogonal slices (transverse, sagittal and coronal) intersecting at the tip of the instrument are computed in the 3D pre-operative CT data.
- "Double-pseudo" representation: two pseudo-transverse and pseudo-sagittal slices including the instrument axis are computed in the CT pre-operative 3D data. A "single-pseudo" (pseudo-transverse) counterpart is also used when displayed on the smart phone.
- "Adjustable pseudo" representation: a single pseudo-transverse slice is computed in the 3D pre-operative CT data. Although the presented information is still defined by the instrument position, its orientation can now also be freely adjusted by the operator around the tool axis or a marked position for a deeper exploration of the data close to the instrument.

In order to evaluate the ability to delegate part of the GUI to the smart phone, we compared four solutions:

- "Standard mode": a triple-ortho or double-pseudo representation is displayed on the master screen only.
- "Double mode": a triple-ortho or double-pseudo is displayed on the master screen and a pseudo-transverse image is displayed on the smart phone; the smart phone view is also added to the master screen.
- "Remote mode": an adjustable-pseudo representation is displayed on the smart phone only. The accelerometers control the orientation of the slice around the tool axis or around a marked position.
- "Distributed mode": a standard double-pseudo representation is displayed on the master screen. At any time the user can record the tool position. Then, while the master screen keeps displaying the standard double-pseudo, the user can navigate through the data around the recorded position using the smart phone which displays the adjustable slice.

### 2.3   Experiments

#### 2.3.1   Experiment n°1

Three experimental conditions were tested: (TO) standard-mode with triple ortho representation of data, (TO+iP) double mode with triple-ortho on the master screen and pseudo-transverse on the smart phone, (TO+iPA) distributed mode with triple-ortho on the master screen and adjustable pseudo-transverse on the smart phone. 30 operators were involved: 12 clinicians and 18 non clinicians (PhD and Master students). Training was performed before the experiment with the set-up presented in section 2.1 and a synthetic CT dataset. After training the dataset was replaced by a real exam of a patient having a quite big and easily detectable renal cyst; the target was the cyst; the user could scan the exam before starting the punctures. The three conditions were presented in a random order to the operator. Speed of execution of the puncture and rate of success are recorded. Between two exercises, a 5mn rest was left to the user. After 10mn of unsuccessful trial, the puncture was considered as a failure. After each experiment the operator had to fill a questionnaire (about his/her fatigue, cognitive effort, liking of the tested solution with Likert scales from 1 to 7) and was eventually asked to give a ranking of the three solutions.

### 2.3.2   Experiment n°2

A second experiment was set where only pseudo-slices were used for representation. Three experimental conditions were tested: (DP) standard mode with double-pseudo representation of data, (iPA) remote mode with an adjustable pseudo-transverse on the smart phone, (DP+iPA) distributed mode with double-pseudo on the master screen and adjustable pseudo-transverse on the smart phone. The incremental nature of the representation involved that the preliminary training is performed in this specific order. But the experiment itself was here again performed with a random order. After the training the dataset is replaced by a real exam of a patient; the target is a simulated hepatic cyst placed in a delicate anatomical area; the user could still scan the exam before starting the punctures. 6 operators (all clinicians) contributed to this experiment. The order of tested conditions, time condition for failure, recorded parameters and questionnaires were similar to experiment n°1. The distance from the tip of the instrument to the target was recorded when the user considers that it had been reached.

## 3      Results

Comparisons of the three conditions in both experiments used a non parametric test (Friedman test) with paired samples (in each experiment, each operator experimented three conditions for the same task). For experiment n°1 where two populations were involved (clinicians and non clinicians), comparisons between the two populations used the Mann-Whitney test.

### 3.1    Experiment n°1

For the global population of 12+18 subjects, the felt comfort, felt cognitive effort and felt fatigue are similar in the three conditions. The duration of targeting for TO+iPA is in average longer than for TO which is longer than TO+iP; only the difference between TO+iPA and TO+iP is statistically significant. The liking of the interaction mode is in average lower for TO than for TO+iP which is lower than for TO+iPA; the liking of TO is significantly different from the other two conditions. Regarding the ranking of the interaction modes, in average TO+iP was preferred to TO+iPA which was preferred to TO; only the difference between TO+iP and TO is statistically significant.

When focusing on the clinician subgroup, all the measured or felt characteristics were similar in the three conditions. When focusing on the non clinician subgroup, the felt comfort, the felt cognitive effort, the felt fatigue and the ranking are similar in the three conditions. The measured duration is in average lower for TO than for TO+iP which is lower than for TO+iPA; the duration for TO+iPA is significantly different from the duration for the other two conditions. In average the liking of TO is lower than the liking for TO+iP which is lower than the liking for TO+iPA; the liking for TO is significantly different than the liking for the other two conditions.

When comparing the two subgroups for the measured duration and for liking of the interaction modes, no difference could be exhibited between the subgroups. This probably means that the difference between statistics for the global population as compared to statistics for the subgroups is due to the number of people involved.

### 3.2     Experiment n°2

Regarding the measured duration, measured accuracy, felt cognitive effort, and felt fatigue, no significant difference was computed among the three conditions. The felt comfort of the iPA condition was significantly different (lower) than the felt comfort of the other two conditions. The liking of the interaction mode was in average lower for iPA than for DP which was lower than DP+iPA; only iPA was significantly different from DP+iPA. Finally, for the ranking of the three types of interaction modes, all the subjects placed the iPA in the last rank; in average DP+iPA was the best placed, before DP and iPA. A significant difference between iPA and DP+iPA could be exhibited.

## 4     Discussion and conclusion

In the first experiment, the increase of duration measured for the TO+iPA condition probably comes from the time spent to explore locally the data with the adjustable slice. Other experiments would be necessary to determine if any particular stage of the puncture is preferably concerned with this adjustment (initial orientation of the needle? fine approach to the target? other?). The fact that TO is in average longer than TO+iP could be explained by the fact that the smart phone with the pseudo-inverse slice adds some useful information that makes the puncture easier with respect to the TO representation alone. However since the difference was not statistically significant this explanation has to be taken with special care and specific additional experiments would be needed.

As concerns the interface distribution on the two displays, the users did not appreciated the remote mode (experiment n°2) where guiding information is only present on the smart phone and they felt uncomfortable with it. The size of the display and the resolution of the displayed data were mentioned by the users as the main limitations. The available zooming function was however nearly never used although systematically introduced.

The combination of displays in a distributed mode was generally appreciated in the conducted experiments. From our point view, the master screen brings good quality information enabling a global 3D perception of the conducted task; however the visualized data directly depend on the position and orientation of the tool. While this is particularly useful for the initial orientation of the needle before entering the tissues, the progress of the needle limits any further exploration of the data. This is probably where the smart phone is the most useful. It allows a local exploration around the current tool position for instance for controlling the absence of anatomical obstacles or the presence of remarkable anatomical features for an easier and safer access to the target.

One limitation of this work is the relatively small number of subjects. The fact that the experiments were rather long due to training, multiple conditions, associated interviews and filling of questionnaires was an obstacle to the recruitment of clinicians even though we moved to the hospital to make their involvement as short and as easy as possible. Having more clinicians would allow classifying them in terms of their expertise of surgical navigation. It would also permit to draw more definitive conclusions about the best representation and interaction modes. Other applications with different type of assistance could also certainly benefit from such integrated technology.

However we think that the presented experiments show that the use of a mini-screen for CAS guidance is feasible, well accepted and is probably a good complement to a larger screen.